\documentstyle[12pt,twoside,cmp209,graphicx,amsmath,amssymb,epsfig]{article}
\newcommand{\be}{\begin{equation}}
\newcommand{\ee}{\end{equation}}

\title[High-temperature series expansions for random Potts models]
      {High-temperature series expansions for random Potts models} 
\author[Hellmund and Janke]
	{Meik Hellmund\refaddr{*}, Wolfhard Janke\refaddr{**}, 
        }
\addresses{
        \addr{*} {Mathematisches Institut, Universit\"at Leipzig,}\\%[-1mm]
        {Augustusplatz 10/11, D-04109 Leipzig, Germany}\\[2mm]
        \addr{**} {Institut f\"ur Theoretische Physik, Universit\"at Leipzig,}\\%[-1mm]
        {Augustusplatz 10/11, D-04109 Leipzig, Germany}\\[2mm]
        {\footnotesize\tt meik.hellmund@math.uni-leipzig.de}\\%[-1mm]
        {\footnotesize\tt wolfhard.janke@itp.uni-leipzig.de}\\[-3mm]
}
\begin{document}

\maketitle

\begin{abstract}
We discuss recently generated high-temperature series expansions for 
the free energy and the susceptibility of random-bond $q$-state Potts 
models on hypercubic lattices. Using the star-graph expansion technique
quenched disorder averages can be calculated exactly for arbitrary 
uncorrelated coupling distributions while keeping 
the disorder strength $p$ as well as the dimension $d$ as symbolic 
parameters. We present analyses of the new series for the susceptibility
of the Ising ($q=2$) and 4-state Potts model in three dimensions up to
order 19 and 18, respectively, and compare our findings with results from
field-theoretical renormalization group studies and Monte Carlo
simulations.

\ {Lattice theory and statistics (Ising, Potts, etc.)},
      {64.60.Fr}\ {Equilibrium properties near critical points,
critical exponents},
      {75.10.Hk}\ {Classical spin models},
      {75.10.Nr}\  {Spin-glass and other random models}
\end{abstract}

\section{Introduction\label{Sec:Intro}}
Systematic series expansions \cite{domb3} for statistical physics models 
defined on a lattice 
provide an useful complement to field-theoretical 
renormalization group studies and large-scale numerical Monte Carlo
simulations. This is in particular true 
when studying phase transitions and critical phenomena 
of quenched, disordered systems. 
In the field-theoretic treatment \cite{folk01} the necessary average over disorder 
realizations at the level of the free energy requires the application of the 
so-called ``replica trick'' which loosely speaking introduces $n$ different, 
interacting copies of the original system, with the formal limit 
$n \rightarrow 0$ taken at the end. In the numerical approach the average over 
a large but finite number of different disorder realizations can, at least in 
principle, be performed explicitly but is very time consuming such that only 
few points in the vast parameter space of the systems can be sampled with 
realistic effort. Moreover extrapolations of the data on finite lattices to 
the infinite-volume limit are required. Using high-temperature series 
expansions, on the other hand, one can obtain for many quantities exact 
results up to a certain order in the inverse temperature. Here the quenched 
disorder is treated exactly and the infinite-volume limit is implicitly 
implied. Moreover, one can keep the disorder strength $p$ as well as the 
dimension $d$ as symbolic parameters and therefore analyse large regions of 
the parameter space of disordered systems. 
The critical part of the series
expansion approach lies in the extrapolation techniques which are used
in order to obtain information on the phase transition behaviour from
the finite number of known coefficients of the high-temperature series.
While for pure systems this usually works quite well,
one can question the use of these extrapolation techniques in disordered
systems, where the singularity structure of the free energy or susceptibility
may be very complicated, involving Griffiths-type singularities or logarithmic
corrections \cite{card99}.

Pure Potts models show either first- or second-order phase transitions,  
depending on the dimension $d$ and the number of states $q$. Since in the 
second-order case the specific-heat exponent $\alpha$ is non-negative
for this class of models, the Harris criterion \cite{harris} suggests for
the corresponding disordered systems either the appearance of a new random 
fixed point ($d=2$, $q=3,4$ and $d=3$, $q=2$) or logarithmic corrections
to the pure fixed point ($d=2$, $q=2$). At first-order phase transitions, 
randomness softens the transitions~\cite{ImryWortis79}. For $d=2$ even 
infinitesimal disorder 
induces a continuous transition~\cite{aiz,hui}, whereas for $d=3$, $q>2$ a 
tricritical point at a finite disorder strength is expected~\cite{card97}.

In this work we studied these scenarios by means of ``star-graph'' 
high-temper\-ature series expansions where the disorder average can be taken 
at the level of individual graphs. Using 
optimized cluster algorithms for the symbolic, exact calculation of 
spin-spin correlators
on finite graphs with arbitrary inhomogeneous couplings, we obtained series 
expansions for the free energy and susceptibility in the inverse temperature
up to order 19 respectively 18
for bond-diluted Ising and Potts 
models in dimensions $d \le 5$, and up to order 17 in arbitrary dimensions. 
Here we shall focus on analyses
of these series in three dimensions where a direct comparison with field-theoretic
renormalization group studies and recent Monte Carlo simulations is possible.

\section{Model}\label{sec:m}

The ferromagnetic disordered $q$-state Potts model on hypercubic lattices 
${\mathbb Z}^d$ is defined by the partition function
\begin{equation}
  \label{eq:1}
  Z = \sum_{\{s_i\}} \exp \left(\beta \sum_{\langle ij\rangle }  
      J_{ij} \delta_{s_i, s_j}\right),
\end{equation}
where $\beta=1/k_B T$ is the inverse temperature, $J_{ij}$ are quenched
(non-negative) nearest-neighbour coupling constants, the spins can take the
values $s_i=1,\ldots,q$, and $\delta_{.,.}$ is the Kronecker symbol.
In our series expansion the combination
\begin{equation}
  \label{eq:v2}
 v_{ij} = \frac{e^{\beta J_{ij}}-1}{e^{\beta J_{ij}}-1+q}
\end{equation}
will be the relevant expansion parameter which in the Ising case ($q=2$) simplifies
to $v_{ij} = \tanh(\beta J_{ij}/2)$. In the symmetric high-temperature phase, the
susceptibility associated with the coupling 
$\sum_i h_i (q \delta_{s_i,1}-1)/(q-1)$ to an external field $h_i$
is given for a  graph with $N$ spins by summing over all two-point correlations,
\begin{equation}
  \label{eq:s1}
  \chi = \frac{1}{N} \sum_i \sum_j \left[ \left\langle \frac{q \delta_{s_i, s_j}
        -1}{q-1}\right\rangle \right]_{\rm av}.
\end{equation}
Here the brackets $[\ldots]_{\rm av}$ indicate
the quenched disorder average which in our case is taken 
over an uncorrelated bimodal distribution of the form
\begin{equation}
  \label{eq:bi}
  P(J_{ij}) = (1-p) \delta(J_{ij}-J_0) + p \delta(J_{ij}- RJ_0).
\end{equation}
Besides bond dilution ($R=0$), which will be in the focus of the present work, 
this also includes random-bond ferromagnets ($0<R<1$) and the physically very
different class of spin glasses ($R=-1$) as special cases. Other distributions 
such as Gaussian distributions can, in principle, also be considered with our 
method.

\section{Series generation methodology}
\label{sec:s}

In this section we briefly review the main technical ingredients necessary for  
our high-temperature series study. We begin with a few basic notations from 
graph theory. A  graph of order $E$  consists of $E$ links connecting $N$ 
vertices. We consider only connected, undirected  graphs that are simple:
no link starts and ends at the same vertex (no tadpoles) and two vertices 
are never connected by more than one link.
Subgraphs are defined by the deletion of links. In this process, isolated
vertices can be dropped. Since each link may be present or absent, a graph 
of order $E$ has $2^E$ (not necessarily non-isomorphic) subgraphs.
These subgraphs may consist of several connected components and are called
clusters. If the deletion of one vertex renders the graph disconnected, such
a vertex is termed articulation point. The ``star graphs'' we are considering 
here are thus just defined by the absence of such articulation points. A graph is 
bipartite if the vertices can be separated into red and black vertices so that
no link connects two vertices of the same color. Equivalently, all closed 
paths in the graph consist of an even number of links. Hypercubic lattices
are evident examples for bipartite graphs.

There are a couple of well-established methods \cite{domb3} known for the 
systematic generation of high-temperature series expansions which differ in
the way relevant subgraphs are selected or grouped together. A recently
developed alternative method \cite{arisue2003} exploits ideas from so-called 
finite-lattice methods, usually employed before for the generation of 
{\em low\/}-temperature series. Using a clever reformulation of the method, 
Arisue {\em et al.\/} \cite{arisue} succeeded to generate a very impressive 32th order 
world-record high-temperature susceptibility series for the pure 
Ising model in three dimensions.

For the class of classical O($N$) spin models without disorder, 
quite long series 
(up to order $\beta^{25}$) have also been produced by linked-cluster 
expansions \cite{butera02}. This technique also allows one to obtain series for 
more involved observables (such as the second moment of the spin-spin 
correlation function yielding the correlation length) which have no star-graph 
expansion. 
Furthermore, it works with free embeddings of graphs into the lattice which 
can be counted orders of magnitude faster than the weak embedding numbers 
needed by the star-graph technique. Nonetheless, the linked-cluster method 
has not yet been applied to problems with quenched disorder.

The star-graph method can be adopted to systems involving quenched disorder
\cite{rap1,singh87} (as also can the no-free-end method \cite{harris-nfe})
 since 
it allows one to 
take the disorder average on the level of individual graphs. The basic idea 
is to assemble the  value of some extensive thermodynamic quantity $F$ on a 
large or even infinite graph from its values on subgraphs: Graphs constitute
a partially ordered set under the ``subgraph'' relation. Therefore, for every
function  $F(G)$ defined on the set of graphs exists another function $W_F(G)$
such that $F(G) = \sum_{g \subseteq  G} W_F(g)$, for all graphs $G$. This 
function can be calculated recursively via 
$W_F(G) = F(G) -  \sum_{g \subset   G} W_F(g)$, resulting for an infinite 
(e.g.\ hypercubic) lattice in 
$F({\mathbb Z}^d) = \sum_G (G:{\mathbb Z}^d)\, W_F(G)$, where 
$(G:{\mathbb Z}^d)$ denotes the weak embedding number of the graph $G$ in the
given lattice structure~\cite{martin74}.

The following observation makes this a useful method: Let $G$ be a graph with
an articulation vertex where two star subgraphs $G_{1,2}$ are glued together.
Then $W_F(G)$ vanishes if $F(G) = F(G_1) + F(G_2)$. An observable $F$ for 
which this property is true on arbitrary graphs with articulation points 
allows a star-graph expansion. All non-star graphs have zero weight $W_F$ in
the sum for $F({\mathbb Z}^d)$. It is easy to see that the (properly 
normalized) free energy $\log Z$ has this property and it can be
proved~\cite{singh87} that the inverse susceptibility $1/\chi$ has it, too, 
even for arbitrary inhomogeneous couplings $J_{ij}$. This restricts the 
summation for $F({\mathbb Z}^d)$ to a sum over star graphs. The linearity of 
the recursion relations then enables the calculation of quenched averages over
the coupling distribution on the level of individual graphs.

The resulting recipe for the susceptibility series is:

\begin{itemize}
\item Graph generation and embedding number counting.
\item Calculation of $Z(G)$ and the correlation matrix\\
$M_{nm}(G) = {\mathrm Tr}\, (q\delta(S_n,S_m)-1) e^{-\beta H(\{J_{ij}\})}$\\
for all graphs as polynomials in $E$ variables $v_{ij}$ defined in (\ref{eq:v2}).
\item Inversion of the $Z$ polynomial as a series up to the desired order.
\item Averaging over quenched disorder,\\
$N_{nm}(G) = \left[ M_{nm}/Z \right]_{P(J)},$\\
  resulting in a matrix of polynomials in $(p,v)$.
\item Inversion of the matrix $N_{nm}$  and subgraph subtraction,\\
 $W_\chi (G) = \sum_{n,m} (N^{-1})_{nm} - \sum_{g\subset G} W_\chi(g)$.
\item Collecting the results from all graphs,\\
$1/\chi = \sum_G (G:{\mathbb Z}^d)\; W_\chi(G)$.
\end{itemize}

\begin{figure}[h]
\begin{center}
  \includegraphics[scale=.45,angle=-90]{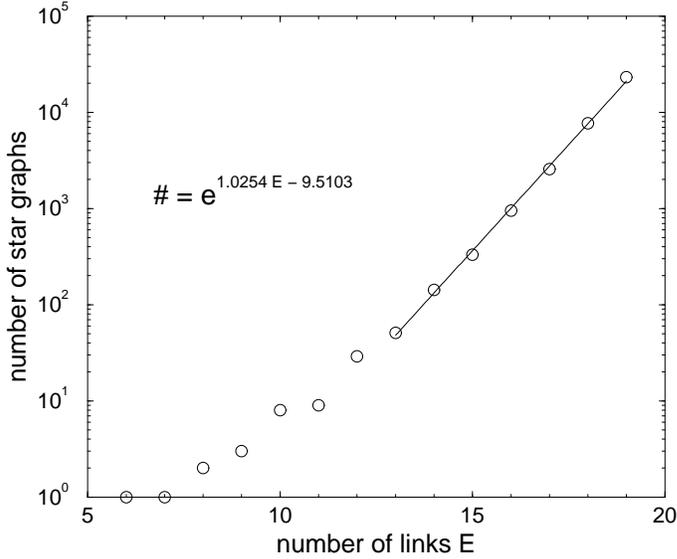}
  \caption{Growth behaviour of the number of  star graphs with $E$ links 
  that can be embedded in hypercubic lattices ${\mathbb Z}^d$.}
  \label{fig:growth}
\end{center}
\end{figure}

Algorithmically the most cumbersome part of this recipe is the first step,
the generation of star graphs and calculation of their (weak) embedding 
numbers. The graph generation is usually done by recursively adding nodes and 
edges to a list of smaller graphs. To make sure that no double counting occurs
this requires an isomorphism test, i.e., the decision whether two
given adjacency lists or adjacency matrices describe the same graph modulo
relabelling and reordering of edges and nodes. We employed the {\tt NAUTY}
package by McKay \cite{mckay81} which allows very fast isomorphism tests by
calculating a canonical representation of the automorphism group of the graphs.
By this means, we
classified for the first time all star graphs up to order 19 that can be
embedded in hypercubic lattices, see Table~\ref{tab:1}. As with any series
expansion, the effort grows exponentially with the maximal order of the 
expansion, rendering each new order roughly as ``expensive'' as all previous 
orders taken together. This is illustrated in Fig.~\ref{fig:growth} where already the
number of star graphs is seen to grow exponentially as a function of the
links $E$. The exponential fit in the range $E=13$ -- 19 suggests that the 
number of star graphs increases roughly by a factor of 2.8 in each of the next higher
orders, predicting about $65\,000$ different star graphs with $E=20$ and about
$180\,000$ with $E=21$.

\begin{table}[t]
\begin{center}
\caption{\label{tab:1}Number of star graphs with $E \ge 8$ links and
non-vanishing embedding numbers on ${\mathbb Z}^d$. For $E=1,4,6$, and 7
only a single star graph exists.}
\vspace*{0.2cm}
  \begin{tabular}{l*{11}{r}r}
   \hline\hline
    order $E$&8&9&10&11&12&13&14&15&16&17&18&19\\
    \hline
    $\#$  &2&3&8&9&29&51&142&330&951&2561&7688&23078\\
   \hline\hline
  \end{tabular}
\end{center}
\end{table}

For each of these graphs we
calculated their (weak) embedding numbers for $d$-dimensional hypercubic
lattices (up to order 17 for arbitrary $d$,  order 18 (general $q$-state
Potts) and 19 (Ising) for dimensions $d \leq 5$). Two typical results are 
depicted in Fig.~\ref{fig:emb}.
For the embedding count we implemented
a refined version of the backtracing algorithm by
Martin \cite{martin74}, making use of a couple of simplifications for 
bipartite hypercubic lattices ${\mathbb Z}^d$. After
extensive tests to find the optimal algorithm
for the ``innermost'' loop, the test for collisions in the embedding, we ended
up using optimized hash tables.

\begin{figure}
\unitlength6mm
\begin{center}
  \begin{picture}(8,1)
\thicklines
\put(0,0){\line(1,0){8}}
\put(0,0){\line(0,1){1}}
\put(0,1){\line(1,0){8}}
\put(8,0){\line(0,1){1}}
\put(3,0){\line(0,1){1}}
\put(0,0){\circle*{.24}}
\put(1,0){\circle*{.24}}
\put(2,0){\circle*{.24}}
\put(3,0){\circle*{.24}}
\put(4,0){\circle*{.24}}
\put(5,0){\circle*{.24}}
\put(6,0){\circle*{.24}}
\put(7,0){\circle*{.24}}
\put(8,0){\circle*{.24}}
\put(0,1){\circle*{.24}}
\put(1,1){\circle*{.24}}
\put(2,1){\circle*{.24}}
\put(3,1){\circle*{.24}}
\put(4,1){\circle*{.24}}
\put(5,1){\circle*{.24}}
\put(6,1){\circle*{.24}}
\put(7,1){\circle*{.24}}
\put(8,1){\circle*{.24}}
\end{picture}
\end{center}
\small
\begin{eqnarray*}
   & 7620 \binom{d}{2} +  76851600    \binom{d}{3}+ 14650620864 \binom{d}{4}\\
 &+\; 404500471680\binom{d}{5}+ 3355519311360   \binom{d}{6}
\end{eqnarray*}

\begin{center}
  \begin{picture}(6,3)
\thicklines
\put(0,0){\line(1,0){3.9}}
\put(1,1){\line(1,0){3.9}}
\put(0,0){\line(1,1){1}}
\put(3.9,0){\line(1,1){1}}
\put(4.9,1){\line(0,1){1.3}}
\put(2.6,0){\line(0,1){1.3}}
\put(1.3,1.3){\line(1,0){2.6}}
\put(2.3,2.3){\line(1,0){2.6}}
\put(1.3,1.3){\line(1,1){1}}
\put(2.6,1.3){\line(1,1){1}}
\put(3.9,1.3){\line(1,1){1}}
\put(0,0){\circle*{.24}}
\put(1,1){\circle*{.24}}
\put(1.3,0){\circle*{.24}}
\put(2.6,0){\circle*{.24}}
\put(3.9,0){\circle*{.24}}
\put(2.3,1){\circle*{.24}}
\put(3.6,1){\circle*{.24}}
\put(4.9,1){\circle*{.24}}
\put(1.3,1.3){\circle*{.24}}
\put(2.6,1.3){\circle*{.24}}
\put(3.9,1.3){\circle*{.24}}
\put(2.3,2.3){\circle*{.24}}
\put(3.6,2.3){\circle*{.24}}
\put(4.9,2.3){\circle*{.24}}
\end{picture}
\end{center}
\small
\begin{eqnarray*}
   &12048 \binom{d}{3}+  396672\binom{d}{4} +  2127360\binom{d}{5}+
2488320\binom{d}{6}
\end{eqnarray*}

  \caption{Two star graphs of order 17 and 19
and their weak embedding numbers up to 6 dimensions.}
  \label{fig:emb}
\end{figure}

The second step of the series generation requires the exact calculation
of the partition function and the matrix of correlations $M_{nm}$ for each 
star graph with arbitrary symbolic couplings $J_{ij}$ defined on the $E \le 19$
edges. The crucial observation is that this can be done most efficiently by
using the cluster representation
\begin{eqnarray}
  \label{eq:cl1}
  Z \propto {\mathcal Z} &=& q^{-N} {\mathrm Tr} \prod_{\langle ij\rangle}\left[1-v_{ij} +v_{ij} q \delta_{s_i,s_j}
    \right]
\nonumber\\
   &=&\!\sum_{C}  q^{e+c-N}\!\! \left(\prod_{\langle ij\rangle \in C}
 v_{ij}\!\right)\!\!\left( \prod_{\langle ij\rangle \notin C} (1-v_{ij})\! \right)\!,\label{eq:cl2}
\end{eqnarray}
where the sum goes over all clusters $C \subseteq G$, $e$ is the number of 
links of the cluster and $c$ the number of connected components of $C$. The
reduced partition function
${\mathcal Z} \equiv Z q^{E-N}/\prod_{\langle ij\rangle} (e^{\beta J_{ij}}-1+q)$ 
is normalized such that $\log \mathcal Z$ has a star-graph
expansion. 
Similarly, the calculation of the susceptibility involves the matrix of correlations
\begin{equation}
  \label{eq:clu}
M_{nm} \propto  \sum_{C_{nm}} q^{e+c-N} \left( \prod_{\langle ij\rangle \in C}
 v_{ij} \right)\left( \prod_{\langle ij\rangle \notin C} (1-v_{ij})\right),
\end{equation}
where the sum is restricted to all clusters $C_{nm}\subseteq G$ in
which the vertices $n$ and $m$ are connected.

This representation essentially reduces the summation over $q^N$
states to a sum over $2^E$ clusters which, compared with previous implementations,
results is a huge saving factor in computing time (of the order of
$10^6$). Further improvements result if the $2^E$ clusters belonging to a graph
are enumerated by Gray codes~\cite{numrec}
such that two consecutive clusters in the sum (\ref{eq:cl2}) differ by exactly
one (added or deleted) link.
In the Ising case $q=2$ another huge
simplification takes place since only clusters where all vertices are of even 
degree contribute to the cluster sum.

Since general purpose software for symbolic manipulations turned out to be too
slow for our purposes, we developed a C\raise2pt\hbox{\tiny++} template 
library using an expanded degree-sparse representation of polynomials and 
series in many variables. For arbi\-trary-precision arithmetics the open source
library GMP was used. Finally, for the case of bond dilution 
($R=0$ in (\ref{eq:bi})) considered here, we made use of the fact that the 
disorder average is most easily calculated via
\begin{equation}
  \label{eq:bd}
 [v_1^{n_1}\ldots v_k^{n_k}]_{\rm av} = (1-p)^k v_0^{n_1+\ldots+n_k}.
\end{equation}

\section{Series analysis: techniques and results}
\label{sec:a}

\subsection{Bond-diluted 3D Ising model}

Disordered magnetic systems belonging to the 3D Ising model universality
class have been studied extensively in 
experiments \cite{Birgeneau0,Birgeneau1,Belanger1} and also
by field theoretical and numerical methods. 
A comprehensive compilation of recent
results can be found in
\cite{folk99}, showing a wide scatter in the critical exponents
 of different groups, presumably due to large crossover effects.

Our high-temperature series expansion for the susceptibility up to order 19
is given with coefficients as polynomials in $p$,
$\chi(v) = \sum_n a_n(p) v^n$ \cite{hj:cpc2002}.
Therefore it should be
well-suited for the method of partial differential approximants~\cite{pda80}
which was successfully used to analyse series with an anisotropy parameter
 describing the crossover between
3D Ising, XY and Heisenberg behaviour \cite{adler97}. But this method was 
unable to give
conclusive results. Therefore we confined ourselves to a
single-parameter series for selected values of $p$.

\begin{figure}
  \begin{center}
\includegraphics[scale=.33,angle=-90]{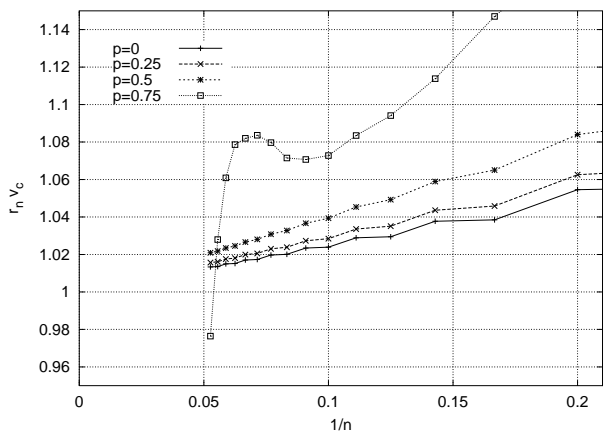}
    \caption{Ratio approximants for different dilutions $p$ vs. $1/n$.
    In order to make them visually comparable, they are (except for $p=0.75$)
     normalized by their respective critical couplings $v_c$.}
    \label{fig:1}
  \end{center}
\end{figure}

The ratio method assumes that the
expected singularity of the form
\begin{equation}
\label{eq:chi_div}
\chi (v) = A (v_c-v)^{-\gamma} + \ldots
\end{equation}
is the closest to the
origin. Then the consecutive ratios of series coefficients  behave
asymptotically as
\begin{equation}
  \label{eq:rat}
  r_n = \frac{a_n}{a_{n-1}} = v_c^{-1} \left(1+\frac{\gamma-1}{n}\right).
\end{equation}
Figure~\ref{fig:1} shows these ratios for different values of $p$.
For small $p$ they show the typical oscillations related to the existence of 
an antiferromagnetic singularity at $-v_c$.  Near the percolation threshold 
at $p_c=0.751\,188$~\cite{lorenz98} (where $T_c$ goes to 0, $v_c$ to 1) the 
series is clearly ill-behaved, related to the $\exp(1/T)$ singularity expected
there. Besides that, the slope (related to $\gamma$) is increasing with $p$.

\begin{table}[t]
   \caption{Transition points $v_c = \tanh(\beta_c J_0/2)$ and critical
exponents $\gamma$ for different dilutions $p$ as obtained from
DLog-Pad{\'e} approximants. \label{tab:2}}
\vspace*{0.2cm}
\addtolength{\tabcolsep}{5pt}
  \begin{center}
    \begin{tabular}{lll}
\hline\hline
\multicolumn{1}{c}{$p$} &
\multicolumn{1}{c}{$v_c$} &
\multicolumn{1}{c}{$\gamma$}\\ \hline
0      & 0.21813(1)  & 1.2493(7)  \\
0.075  & 0.23633(1)  & 1.2589(8)  \\
0.15   & 0.25788(1)  & 1.2714(8)  \\
0.225  & 0.28382(1)  & 1.2873(10) \\
0.3    & 0.31566(2)  & 1.305(4)   \\
0.375  & 0.35557(5)  & 1.329(4)   \\
0.45   & 0.40743(10) & 1.365(6)   \\
0.525  & 0.4772(2)   & 1.400(10)  \\
0.6    & 0.576(1)    & 1.435(60)  \\
\hline\hline
    \end{tabular}
  \end{center}
\end{table}

\begin{figure}[bh]
\begin{center}
\includegraphics[scale=.379,angle=-90]{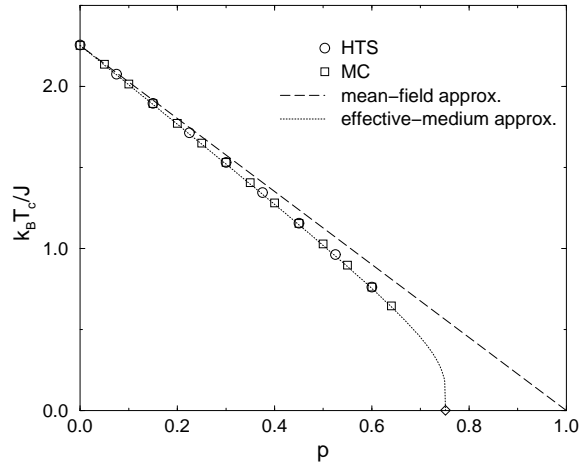}
\caption{Transition temperatures of the bond-diluted Ising model for different 
         dilutions $p$ as obtained from
         our  DLog-Pad{\'e} high-temperature series (HTS) analyses and from
        Monte Carlo (MC) simulations \cite{berche04}. 
	For comparison also the (properly normalized) 
	mean-field and effective-medium approximations are shown.}
\label{fig:Is_Tc}
\end{center}
\end{figure}

The widely used DLog-Pad{\'e} method consists in calculating Pad{\'e}
approximants to
the logarithmic derivative of $\chi(v)$,
\begin{equation}
\frac{d \ln \chi(v)}{d v} = \frac{\gamma}{v_c-v} + \dots.
\label{eq:pade}
\end{equation}
The smallest real pole of the
approximant is an estimation of $v_c$ and its residue gives $\gamma$.
The results presented in Table~\ref{tab:2} are the averages of 45 -- 55 
different Pad{\'e} approximants for each value of $p$, with the error in 
parentheses indicating the standard deviation. The scattering of the Pad{\'e} 
approximants increases with $p$,
getting again inconclusive near the percolation threshold. 
Nevertheless, up to about $p=0.6$ the series estimates for $v_c$
respectively $T_c$ are in perfect agreement\footnote{%
Notice that ``$p$'' in the present notation corresponds to ``$1-p$'' in
Ref.~\cite{berche04}.}
with the Monte Carlo (MC)
results of Ref.~\cite{berche04}. This is demonstrated in Fig.~\ref{fig:Is_Tc}
where also the (properly normalized) mean-field and effective-medium
approximation \cite{Turban80} are shown for comparison.

The critical exponent $\gamma$,
as provided by this method, apparently varies with the disorder strength.
More sophisticated analysis methods,
such as inhomogeneous differential approximants \cite{fisher79,guttmann89}, 
the Baker-Hunter method \cite{bakerhunter} or 
the methods M1 and M2 \cite{adler91}, especially tailored to deal with
confluent singularities as one would
expect in a crossover situation, 
 give improved results in the pure ($p=0$)
case but do not essentially change the results in the presence of disorder.

Thus, while for theoretical reasons we still find it likely that the 
variation of $\gamma$ with the disorder strength can be attributed to 
neglected or insufficiently treated correction terms, it proved
clearly impossible to verify this effect in the series analysis.
In fact, a plot of $\gamma$ vs. $p$ does not even show an indication of
a plateau.
In the central 
disorder regime, $p=0.3$ -- 0.5, the high-temperature
series estimates given in Table~\ref{tab:2} are at least compatible
with Monte Carlo results for site and bond 
dilution \cite{berche04,Calabrese1,balles98}
which cluster quite
sharply around $\gamma_{\rm MC} = 1.34(1)$. Field-theoretic renormalization group
estimates \cite{folk99,Pelissetto1} favor slightly smaller exponents 
of $\gamma_{\rm RG} = 1.32$ -- 1.33,
while experiments \cite{Birgeneau0,Birgeneau1,Belanger1} 
report values between $\gamma_{\rm exp} = 1.31$ -- 1.44, cp.,
e.g., the table in Ref.~\cite{MC_here}.

\subsection{Bond-diluted 4-state Potts model}

In three dimensions the 4-state Potts model exhibits in the pure case a strong
first-order transition \cite{kappler} which is expected to stay first order up to some
finite disorder strength, before it gets softened to a second-order transition
governed by a disorder fixed point.

\begin{figure}[h]
  \begin{center}
\includegraphics[scale=.33,angle=-90]{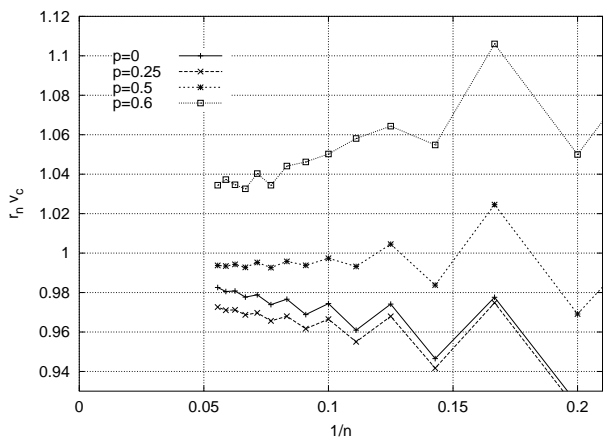}
   \caption{Ratio approximants for different dilutions $p$ vs.\ $1/n$
            (normalized by $v_c$ as in  Fig.~\ref{fig:1}).}
    \label{fig:potts_1}
  \end{center}
\end{figure}

In the latter regime we
are interested in locating power-law divergences of the form (\ref{eq:chi_div}) 
from our susceptibility series up to order 18 \cite{hj:npb2002,hj:pre2003}.
To localize a first-order transition point, however,
a high-temperature series alone is not sufficient since there
the correlation length remains finite and no critical singularity occurs.
In analysing series by ratio, Pad{\'e} or differential approximants, the
approximant will
provide an analytic continuation  of the thermodynamic quantities beyond the
transition point into a metastable region on a pseudo-spinodal line
with a singularity $T^*_c < T_c$ and effective ``critical exponents''
at $T^*_c$.
Again we first employed the ratio method which is the least sophisticated method of 
series analysis, but usually it is quite robust and
gives a good first estimate of the series behaviour.
Figure~\ref{fig:potts_1} shows these ratios for different values of $p$. They 
behave qualitatively similar to the Ising model case (oscillations
caused by the antiferromagnetic singularity at $-v_c$, strong influence
of the percolation point at $p_c \approx 0.75$).
Notice that the slope
($\propto \gamma-1$) is increasing with $p$, changing from $\gamma<1$ to
$\gamma>1$ around $p=0.5$.

\begin{figure}
\begin{center}
\includegraphics[scale=.33,angle=-90]{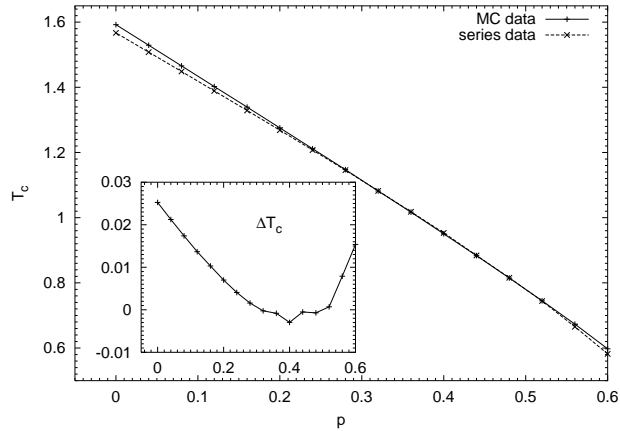}
    \caption{Transition temperatures of the bond-diluted 4-state Potts model 
    for different dilution $p$ as obtained from
    Monte Carlo (MC) simulations \cite{chat01a} and DLog-Pad{\'e} series analyses.
    The inset shows the difference between the two estimates.}
    \label{fig:2}
\end{center}
\end{figure}

Figure~\ref{fig:2} compares the critical temperature, estimated from an
\begin{figure}
\vspace*{-0.4cm}
\begin{center}
\includegraphics[scale=.33,angle=-90]{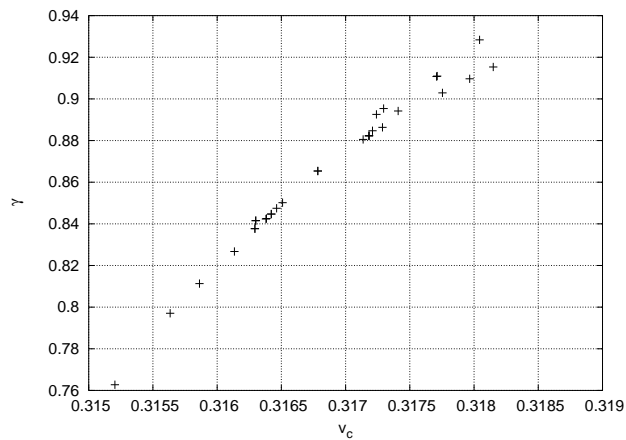}
    \caption{Scattering of different Pad{\'e} approximants at dilution $p=0.4$:
critical exponent $\gamma$ against critical coupling $v_c$.}
    \label{fig:3}
\end{center}
\end{figure}
average of 25 -- 30 Pad{\'e} approximants for each value of $p$,\footnote{%
Again, ``$p$'' in the present notation corresponds to ``$1-p$'' in
Ref.~\cite{chat01a}.}
with the results  of recent Monte Carlo simulations \cite{chat01a}.
For small $p$, in the first-order region, the series underestimates the
critical temperature. As explained above, this is an estimate not of $T_c$
but of $T^*_c$. Between $p=0.3$ and $p=0.5$, the estimates confirm,
within errors, the Monte Carlo results, indicating that now both methods see
the same second-order transition. Beyond $p=0.5$, the scatter of different
Pad{\'e} approximants increases rapidly, related to the crossover to the
percolation point.

\begin{figure}
\includegraphics[scale=.295,angle=-90]{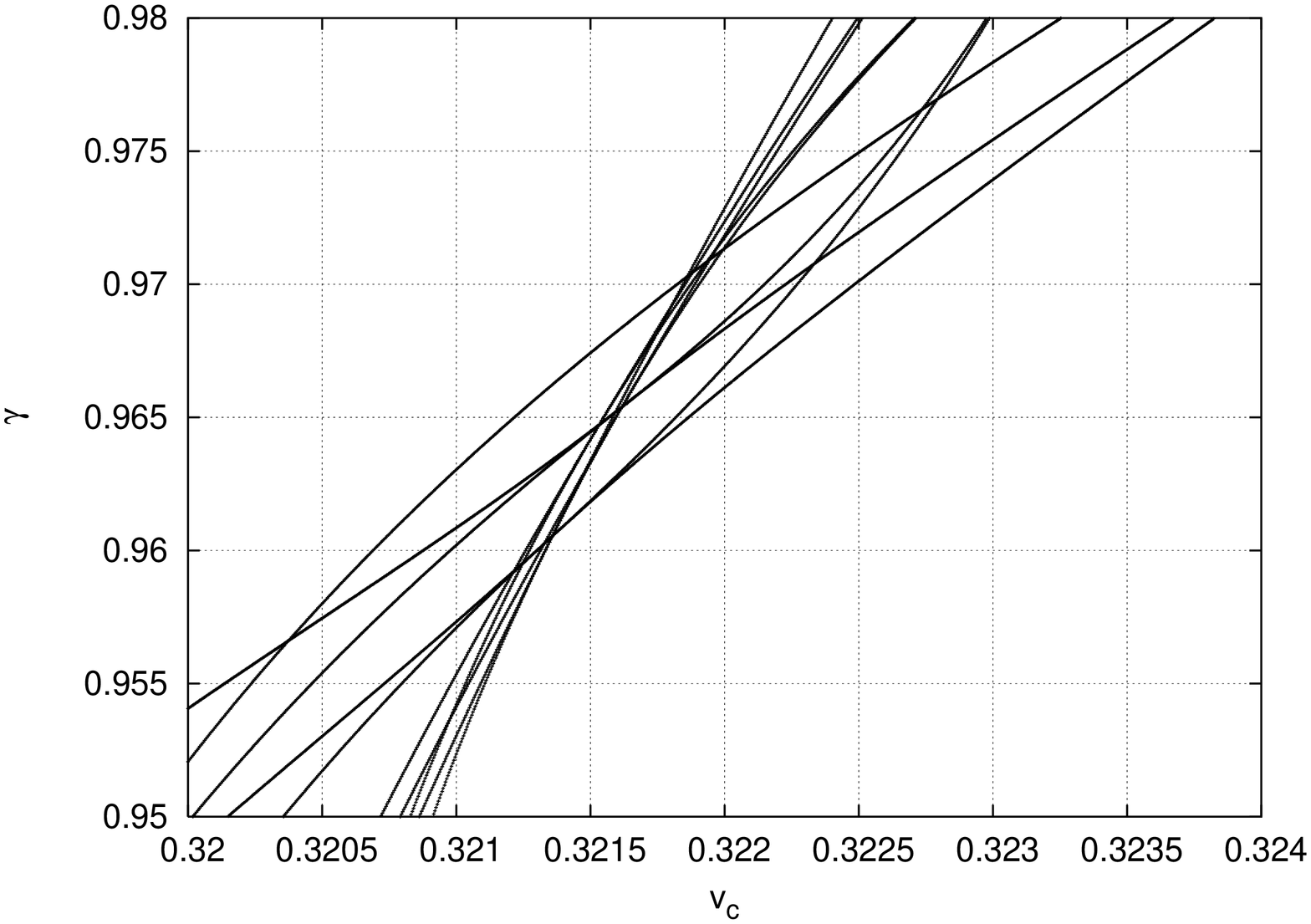}\includegraphics[scale=.295,angle=-90]{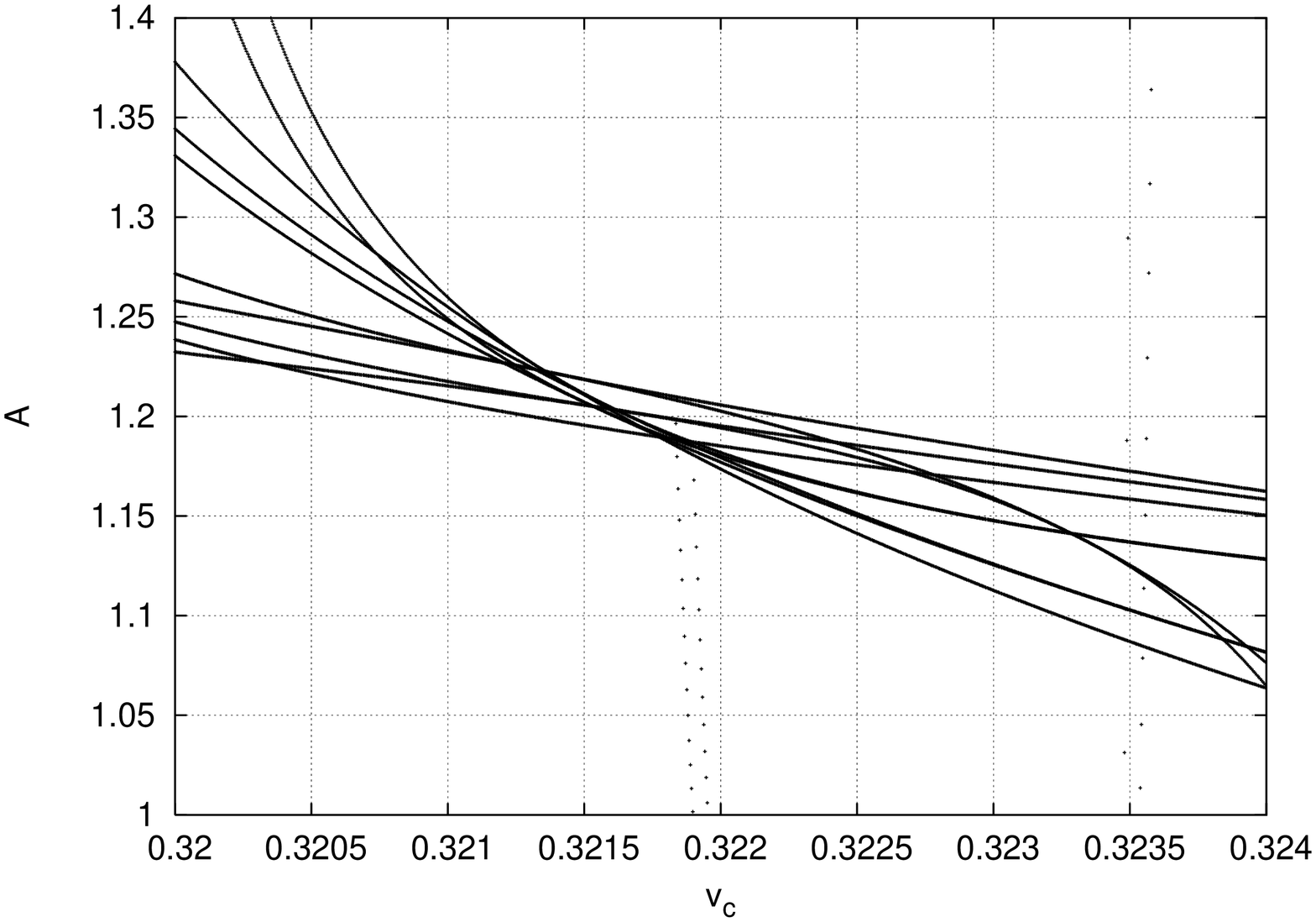}

    \caption{Values for the critical exponent $\gamma$ and  amplitude $A$
at $p=0.4$ as function of trial $v_c$ estimates from the Baker-Hunter analysis.
From the clustering of different Pad{\'e} approximants in both pictures
we estimate  $v_c=0.3217,$ $\gamma=0.966,$ and $A = 1.21$.}
    \label{fig:4}
\end{figure}

The situation is more complicated with respect to the critical exponent $\gamma$.
The DLog-Pad{\'e} analysis gives inconclusive results due to a large scattering
between different Pad{\'e} approximants, as shown in Fig.~\ref{fig:3}. One
possible reason for this failure is the existence of confluent singularities.
The dots in Eq.~(\ref{eq:chi_div}) indicate correction terms which can be
parametrized as follows:
\begin{equation}
  \label{eq:div2}
  \chi (v) = A (v_c-v)^{-\gamma} [1 + A_1(v_c-v)^{\Delta_1} + A_2 (v_c-v)^{\Delta_2} + \ldots],
\end{equation}
where $\Delta_i$ are the confluent correction exponents.
Among the various
sophisticated analysis methods
(inhomogeneous differential approximants~\cite{fisher79,guttmann89}
and the methods M1 and M2~\cite{adler91}),
in the case at hand,
the Baker-Hunter method~\cite{bakerhunter} appeared to be the most successful,
giving consistent results at larger dilutions $p>0.35$ where the leading-term 
DLog-Pad{\'e} analysis failed. The Baker-Hunter method
assumes that the function under investigation has confluent singularities
 \begin{equation}
   \label{eq:b1}
   F(z) = \sum_{i=1}^N A_i \left(1-\frac{z}{z_c}\right)^{-\lambda_i} = \sum_{n=0} a_n z^n ,
 \end{equation}
which can be transformed into an auxiliary function $g(t)$ that is
meromorphic and therefore suitable for Pad{\'e} approximation. After the
substitution $z = z_c(1-e^{-t})$ we expand $F(z(t)) = \sum_n c_n t^n $
and construct the  new series
\begin{equation}
  \label{eq:b2}
g(t) = \sum_{n=0} n!\;c_n\; t^n = \sum_{i=1}^N \frac{A_i}{1-\lambda_it},
\end{equation}
such that Pad{\'e} approximants to $g(t)$ exhibit poles at $t=1/\lambda_i$ with
residues $-A_i/\lambda_i$.
This method is applied by plotting these poles and residues for different
Pad{\'e} approximants to $g(t)$ as functions of $z_c$. The optimal set of values
for the parameters is determined visually from the best clustering of
different Pad{\'e} approximants, as demonstrated in Fig.~\ref{fig:4}.

\begin{figure}
\begin{center}
\includegraphics[scale=.32,angle=-90]{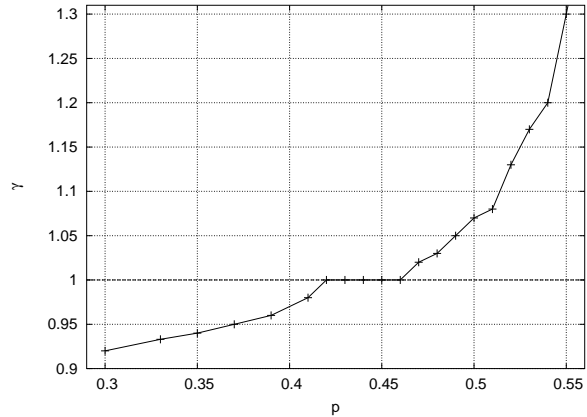}

\caption{Effective critical exponent  $\gamma$ as function of the dilution $p$
from Baker-Hunter analyses.}
  \label{fig:6}
\end{center}
\end{figure}

Using this method, our results for the critical exponent $\gamma$ are plotted in
Fig.~\ref{fig:6}. They show an effective exponent monotonically increasing
with $p$ but reaching a
plateau at $\gamma=1$ for dilutions between
$p=0.42$ and $p=0.46$. The following sharp increase is to be interpreted as
due to the crossover to the percolation fixed point at $p_c \approx 0.75$, $T_c=0$,
where a $\chi \sim \exp(1/T)$ behaviour is expected.

It is well known (see, e.g., Ref.~\cite{barma85})
that series analysis in crossover
situations is extremely difficult. If the parameter $p$ interpolates between
regions governed by different fixed points, the  exponent obtained from a
 finite number of terms of a series expansion must cross somehow
between its  universal values, and does this usually quite slowly.
Therefore it does not come as a
 surprise that the Monte Carlo simulations quoted above see
the onset of a second order phase transition already for smaller values of
the disorder strength $p$.
 The mere existence of a plateau in $\gamma_{\rm eff}(p)$, however,
is an indication that here  truly critical behaviour is seen.
It is  governed by a
fixed point for which we obtain  $\gamma=1.00(3)$.
Here, as always in series analyses, the error estimates the
scattering of different approximants.

\section{Discussion}
\label{sec:c}

We have implemented a comprehensive toolbox for generating and enumerating
star graphs as required for high-temperature series expansions of quenched,
disordered systems.
Monte Carlo simulations of systems with quenched disorder require an enormous
amount of computing time because many realizations have to be simulated for
the quenched average. For this reason it is hardly possible to scan a whole
parameter range. Using high-temperature series expansions, on the other hand,
one can obtain this average exactly. Since the relevant parameters (degree
of disorder $p$, spatial dimension $d$, number of states $q$, etc.)
can be kept
as symbolic variables, the number of potential applications is very large.

Here we presented analyses of the susceptibility series for the three-dimensional 
bond-diluted Ising and 4-state Potts model. The resulting phase diagrams 
in the $p$-$T$-plane are in very good agreement with recent Monte Carlo results.
As far as the critical exponent $\gamma$ is concerned, however, large
crossover effects render a reliable determination from series expansions up to
order 19 respectively 18 very difficult. In the Ising case we estimate values
that are clearly different from the pure case but exhibit a pronounced
dependence on the degree of dilution. For the 4-state Potts model with its
strong first-order phase transition in the pure case, the singularity structure
of the disordered model is even more involved. Still, by comparing the series 
expansions with numerical data we can identify signals for the onset of a 
softening to a second-order transition at a finite disorder strength.

\section*{Acknowledgments}
It is a great pleasure to thank Yurko Holovatch for giving us the opportunity 
to contribute to the Festschrift dedicated to the 60th birthday of Reinhard Folk. 
    Support by  DFG grant No.~JA 483/17-3 and partial support from the
  German-Israel-Foundation under
          grant No.~I-653-181.14/1999 is gratefully acknowledged.

%\twocolumngrid
%\bibliography{star}

\begin{thebibliography}{21}
\expandafter\ifx\csname natexlab\endcsname\relax\def\natexlab#1{#1}\fi
\expandafter\ifx\csname bibnamefont\endcsname\relax
  \def\bibnamefont#1{#1}\fi
\expandafter\ifx\csname bibfnamefont\endcsname\relax
  \def\bibfnamefont#1{#1}\fi
\expandafter\ifx\csname citenamefont\endcsname\relax
  \def\citenamefont#1{#1}\fi
\expandafter\ifx\csname url\endcsname\relax
  \def\url#1{\texttt{#1}}\fi
\expandafter\ifx\csname urlprefix\endcsname\relax\def\urlprefix{URL }\fi
\providecommand{\bibinfo}[2]{#2}
% \providecommand{\eprint}[2][]{\url{#2}}
\providecommand{\eprint}[1]{\url{#1}}




\bibitem{domb3}
\bibinfo{editor}{\bibfnamefont{C.}~\bibnamefont{Domb}} \bibnamefont{and}
  \bibinfo{editor}{\bibfnamefont{M.~S.} \bibnamefont{Green}}, eds.,
  \emph{\bibinfo{title}{Series Expansions for Lattice Models}\/},
  vol.~\bibinfo{volume}{3} of \emph{\bibinfo{series}{Phase Transitions and
  Critical Phenomena}} (\bibinfo{publisher}{Academic Press},
\bibinfo{address}{New York},  \bibinfo{year}{1974}).


%Weak quenched disorder and criticality: resummation of asymptotic(?) series

\bibitem{folk01}
\bibinfo{author}{\bibfnamefont{Y.}~\bibnamefont{Holovatch}},
  \bibinfo{author}{\bibfnamefont{V.}~\bibnamefont{Blavats'ka}},
  \bibinfo{author}{\bibfnamefont{M.}~\bibnamefont{Dudka}},
  \bibinfo{author}{\bibfnamefont{C.}~\bibnamefont{von Ferber}},
  \bibinfo{author}{\bibfnamefont{R.}~\bibnamefont{Folk}}, \bibnamefont{and}
  \bibinfo{author}{\bibfnamefont{T.}~\bibnamefont{Yavors'kii}},
  \emph{\bibinfo{title}{Weak quenched disorder and criticality: Resummation of 
  asymptotic(?) series}\/},
  \bibinfo{journal}{Int. J. Mod. Phys. B} \textbf{\bibinfo{volume}{16}},
  \bibinfo{pages}{4027} (\bibinfo{year}{2002})
  [\eprint{cond-mat/0111158}].


%Logarithmic Correlations in Quenched Random Magnets and Polymers

\bibitem{card99}
\bibinfo{author}{\bibfnamefont{J.}~\bibnamefont{Cardy}},
\emph{\bibinfo{title}{Logarithmic correlations in quenched random magnets and 
polymers}\/},
  preprint \eprint{cond-mat/9911024}.




%Effect of random defects on the critical behaviour of Ising models

\bibitem{harris}
\bibinfo{author}{\bibfnamefont{A.~B.} \bibnamefont{Harris}},
\emph{\bibinfo{title}{Effect of random defects on the critical behaviour of 
Ising models}\/},
  \bibinfo{journal}{Journal of Physics C: Solid State Physics}
  \textbf{\bibinfo{volume}{7}}, \bibinfo{pages}{1671} (\bibinfo{year}{1974}).


\bibitem{ImryWortis79}
Y. Imry and M. Wortis, 
{\em Influence of quenched impurities on first-order phase transitions\/},
Phys. Rev. B {\bf 19}, 3580 (1979).


%Rounding of first-order phase transitions in systems with 
%                  quenched disorder

\bibitem{aiz}
\bibinfo{author}{\bibfnamefont{M.}~\bibnamefont{Aizenman}} \bibnamefont{and}
  \bibinfo{author}{\bibfnamefont{J.}~\bibnamefont{Wehr}},
  \emph{\bibinfo{title}{Rounding of first-order phase transitions in systems with 
  quenched disorder}\/},
  \bibinfo{journal}{Phys. Rev. Lett.} \textbf{\bibinfo{volume}{62}},
  \bibinfo{pages}{2503} (\bibinfo{year}{1989}).


\bibitem{hui}
K.~Hui and A.~N.~Berker, 
{\em Random-field mechanism in random-bond multicritical systems\/},
Phys. Rev. Lett. {\bf 62}, 2507 (1989).


%Critical behaviour of random bond Potts models

\bibitem{card97}
\bibinfo{author}{\bibfnamefont{J.}~\bibnamefont{Cardy}} \bibnamefont{and}
  \bibinfo{author}{\bibfnamefont{J.~L.} \bibnamefont{Jacobsen}},
  \emph{\bibinfo{title}{Critical behaviour of random-bond Potts models}\/},
  \bibinfo{journal}{Phys. Rev. Lett.} \textbf{\bibinfo{volume}{79}},
  \bibinfo{pages}{4063} (\bibinfo{year}{1997}).



\bibitem{arisue2003}
H. Arisue and T. Fujiwara, 
{\em Algorithm of the finite-lattice method for high-temperature expansion 
of the Ising model in three dimensions\/},
Phys. Rev. E {\bf 67}, 066109 (2003).



%Higher orders of the high-temperature expansion for the
%                  Ising model in  three dimensions,

\bibitem{arisue}
\bibinfo{author}{\bibfnamefont{H.}~\bibnamefont{Arisue}},
  \bibinfo{author}{\bibfnamefont{T.}~\bibnamefont{Fujiwara}}, \bibnamefont{and}
  \bibinfo{author}{\bibfnamefont{K.}~\bibnamefont{Tabata}},
  \emph{\bibinfo{title}{Higher orders of the high-temperature expansion for the
  Ising model in three dimensions}\/},
  \bibinfo{journal}{Nucl. Phys. B (Proc. Suppl.)} \textbf{\bibinfo{volume}{129}},
  \bibinfo{pages}{774} (\bibinfo{year}{2004}) [\eprint{hep-lat/0309158}].




%A library of extended high-temperature expansions of basic observables 
%for the spin S Ising models on two- and three-dimensional lattices 

\bibitem{butera02}
\bibinfo{author}{\bibfnamefont{P.}~\bibnamefont{Butera}} \bibnamefont{and}
  \bibinfo{author}{\bibfnamefont{M.}~\bibnamefont{Comi}}, 
  \emph{\bibinfo{title}{A library of extended high-temperature expansions of 
  basic observables for the spin S Ising models on two- and three-dimensional 
  lattices}\/},
  \bibinfo{journal}{J.
  Stat. Phys.} \textbf{\bibinfo{volume}{109}}, \bibinfo{pages}{311}
  (\bibinfo{year}{2002}) [\eprint{hep-lat/0204007}].



%The Ising ferromagnet with impurities: a series expansion approach

\bibitem{rap1}
\bibinfo{author}{\bibfnamefont{D.~C.} \bibnamefont{Rapaport}},
\emph{\bibinfo{title}{The Ising ferromagnet with impurities: A series expansion 
approach. I and II}\/},
  \bibinfo{journal}{Journal of Physics C: Solid State Physics}
  \textbf{\bibinfo{volume}{5}}, \bibinfo{pages}{1830, 2813}
  (\bibinfo{year}{1972}).




%High-temperature series expansion for spin glasses.
%                  {I}. Derivation of the series

\bibitem{singh87}
\bibinfo{author}{\bibfnamefont{R.~R.~P.} \bibnamefont{Singh}} \bibnamefont{and}
  \bibinfo{author}{\bibfnamefont{S.}~\bibnamefont{Chakravarty}},
  \emph{\bibinfo{title}{High-temperature series expansion for spin glasses.
  {I}. Derivation of the series}\/},
  \bibinfo{journal}{Phys. Rev. B} \textbf{\bibinfo{volume}{36}},
  \bibinfo{pages}{546} (\bibinfo{year}{1987}).



%Renormalized (1 / sigma ) expansion for lattice animals and localization

\bibitem{harris-nfe}
\bibinfo{author}{\bibfnamefont{A.~B.} \bibnamefont{Harris}},
\emph{\bibinfo{title}{Renormalized ($1/\sigma$) expansion for lattice animals and 
localization}\/},
  \bibinfo{journal}{Phys. Rev. B} \textbf{\bibinfo{volume}{26}},
  \bibinfo{pages}{337} (\bibinfo{year}{1982}).




%Computer Techniques for Evaluating Lattice Constants

%\bibitem[{\citenamefont{Martin}(1974)}]{martin74}
\bibitem{martin74}
\bibinfo{author}{\bibfnamefont{J.~L.} \bibnamefont{Martin}}, 
\emph{\bibinfo{title}{Computer techniques for evaluating lattice constants}\/},
in  \cite{domb3},
  pp. \bibinfo{pages}{97--112}.



%Practical graph isomorphism

\bibitem{mckay81}
\bibinfo{author}{\bibfnamefont{B.~D.} \bibnamefont{McKay}},
\emph{\bibinfo{title}{Practical graph isomorphism}\/},
  \bibinfo{journal}{Congressus Numerantium} \textbf{\bibinfo{volume}{30}},
  \bibinfo{pages}{45} (\bibinfo{year}{1981}).%,
%  \bibinfo{note}{http://cs.anu.edu.au/{$\sim$}bdm/nauty/}.


%\bibitem[{\citenamefont{Press et~al.}(1992)\citenamefont{Press, Teukolsky,
%  Vetterling, and Flannery}}]{numrec}
\bibitem{numrec}
\bibinfo{author}{\bibfnamefont{W.~H.} \bibnamefont{Press}},
  \bibinfo{author}{\bibfnamefont{S.~A.} \bibnamefont{Teukolsky}},
  \bibinfo{author}{\bibfnamefont{W.~T.} \bibnamefont{Vetterling}},
  \bibnamefont{and} \bibinfo{author}{\bibfnamefont{B.~P.}
  \bibnamefont{Flannery}}, \emph{\bibinfo{title}{Numerical Recipes in C}}
  (\bibinfo{publisher}{Cambridge University Press},
\bibinfo{address}{Cambridge},
\bibinfo{year}{1992}).


\bibitem{Birgeneau0}
R.~J.~Birgeneau, R.~A.~Cowley, G.~Shirane,
H.~Yoshizawa, D.~P.~Belanger, A.~R.~King, and V.~Jaccarino, 
{\em Critical behavior of a site-diluted three-dimensional Ising magnet\/},
Phys. Rev. B {\bf 27}, 6747 (1983).

\bibitem{Birgeneau1}
P.~W.~Mitchell, R.~A.~Cowley, H.~Yoshizawa, 
P.~B\"oni, Y.~J.~Uemura, and R.~J.~Birgeneau,
{\em Critical behavior of the three-dimensional site-random Ising magnet: 
Mn$_{\rm x}$Zn$_{\rm 1-x}$F$_2$\/},
Phys. Rev. B  {\bf 34}, 4719 (1986).

\bibitem{Belanger1}
D.~P.~Belanger, A.~R.~King, and V.~Jaccarino,
{\em Crossover from random-exchange to random-field critical behavior in 
Fe$_{\rm x}$Zn$_{\rm 1-x}$F$_2$\/},
Phys. Rev. B {\bf 34}, 452 (1986).
		
\bibitem{folk99}
R.~Folk, Y.~Holovatch, and T.~Yavors'kii, {\em Effective and asymptotic
critical exponents of weakly diluted quenched {I}sing model: 3d approach
versus $\sqrt{\epsilon}$-expansion\/},
Phys. Rev. B {\bf 61}, 15114 (2000).
% --15129.



%High-temperature series expansions for random-bond Potts models on $Z^d$

\bibitem{hj:cpc2002}
\bibinfo{author}{\bibfnamefont{M.}~\bibnamefont{Hellmund}} \bibnamefont{and}
  \bibinfo{author}{\bibfnamefont{W.}~\bibnamefont{Janke}},
  \emph{\bibinfo{title}{High-temperature series expansions for random-bond 
  Potts models on ${\mathbb Z}^d$}\/},
  \bibinfo{journal}{Comp. Phys. Comm.} \textbf{\bibinfo{volume}{147}},
  \bibinfo{pages}{435} (\bibinfo{year}{2002}).




%Bicriticality and partial differential approximants

\bibitem{pda80}
\bibinfo{author}{\bibfnamefont{M.~E.} \bibnamefont{Fisher}} \bibnamefont{and}
  \bibinfo{author}{\bibfnamefont{J.-H.} \bibnamefont{Chen}}, 
  \emph{\bibinfo{title}{Bicriticality and partial differential approximants}\/},
  in
  \emph{\bibinfo{booktitle}{Phase Transitions: Carg{\`e}se 1980}}, edited by
  \bibinfo{editor}{\bibfnamefont{M.}~\bibnamefont{L{\'e}vy}},
  \bibinfo{editor}{\bibfnamefont{J.~C.} \bibnamefont{Le~Guillou}},
  \bibnamefont{and}
  \bibinfo{editor}{\bibfnamefont{J.}~\bibnamefont{Zinn-Justin}}
  (\bibinfo{publisher}{Plenum}, \bibinfo{address}{New York},
  \bibinfo{year}{1982}), pp. \bibinfo{pages}{169--216}.



%Series analysis of tricritical behaviour: 
%          mean-field model and partial differential approximants

\bibitem{adler97}
\bibinfo{author}{\bibfnamefont{Z.}~\bibnamefont{Salman}} \bibnamefont{and}
  \bibinfo{author}{\bibfnamefont{J.}~\bibnamefont{Adler}},
  \emph{\bibinfo{title}{Series analysis of tricritical behaviour: 
  Mean-field model and partial differential approximants}\/},
  \bibinfo{journal}{Journal of Physics A: Mathematical and General}
  \textbf{\bibinfo{volume}{30}}, \bibinfo{pages}{1979} (\bibinfo{year}{1997}).



%Precise determination of the bond percolation thresholds
%                   and finite-size scaling corrections for the sc, 
%                  fcc, and bcc lattices

\bibitem{lorenz98}
\bibinfo{author}{\bibfnamefont{C.~D.} \bibnamefont{Lorenz}} \bibnamefont{and}
  \bibinfo{author}{\bibfnamefont{R.~M.} \bibnamefont{Ziff}},
  \emph{\bibinfo{title}{Precise determination of the bond percolation thresholds
  and finite-size scaling corrections for the sc, 
  fcc, and bcc lattices}\/},
  \bibinfo{journal}{Phys. Rev. E} \textbf{\bibinfo{volume}{57}},
  \bibinfo{pages}{230} (\bibinfo{year}{1998}).


%Bond dilution in the 3D Ising model: a Monte Carlo study

\bibitem{berche04}
\bibinfo{author}{\bibfnamefont{P.~E.} \bibnamefont{Berche}},
  \bibinfo{author}{\bibfnamefont{C.}~\bibnamefont{Chatelain}},
  \bibinfo{author}{\bibfnamefont{B.}~\bibnamefont{Berche}}, \bibnamefont{and}
  \bibinfo{author}{\bibfnamefont{W.}~\bibnamefont{Janke}},
  \emph{\bibinfo{title}{Bond dilution in the 3D Ising model: A Monte Carlo study}\/},
  \bibinfo{journal}{Eur. Phys. J. B} \textbf{\bibinfo{volume}{38}},
  \bibinfo{pages}{463} (\bibinfo{year}{2004}).


\bibitem{Turban80}
L. Turban, 
{\em Effective-medium approximation for quenched bond-disorder in the Ising model\/},  
Phys. Lett. A {\bf 75}, 307 (1980).


%Inhomogeneous differential approximants for power series

\bibitem{fisher79}
\bibinfo{author}{\bibfnamefont{M.~E.} \bibnamefont{Fisher}} \bibnamefont{and}
  \bibinfo{author}{\bibfnamefont{H.}~\bibnamefont{Au-Yang}},
  \emph{\bibinfo{title}{Inhomogeneous differential approximants for power series}\/},
  \bibinfo{journal}{Journal of Physics A: Mathematical and General}
  \textbf{\bibinfo{volume}{12}}, \bibinfo{pages}{1677} (\bibinfo{year}{1979}).


%Asymptotic Analysis of Power-Series Expansions

\bibitem{guttmann89}
\bibinfo{author}{\bibfnamefont{A.~J.} \bibnamefont{Guttmann}}, 
\emph{\bibinfo{title}{Asymptotic analysis of power-series expansions}\/},
in
 vol.~\bibinfo{volume}{13} of
  \emph{\bibinfo{series}{Phase Transitions and Critical Phenomena}},
edited by
\bibinfo{editor}{\bibfnamefont{C.}~\bibnamefont{Domb}} \bibnamefont{and}
  \bibinfo{editor}{\bibfnamefont{J.~L.} \bibnamefont{Lebowitz}}
  (\bibinfo{publisher}{Academic Press}, \bibinfo{address}{New York},
  \bibinfo{year}{1989}), pp. \bibinfo{pages}{1--234}.



%Methods of series analysis II

\bibitem{bakerhunter}
\bibinfo{author}{\bibfnamefont{G.~A.}~\bibnamefont{Baker}}
  \bibnamefont{and} \bibinfo{author}{\bibfnamefont{D.~L.}
  \bibnamefont{Hunter}}, 
  \emph{\bibinfo{title}{Methods of series analysis II}\/},
  \bibinfo{journal}{Phys. Rev. B}
  \textbf{\bibinfo{volume}{7}}, \bibinfo{pages}{3377} (\bibinfo{year}{1973}).


%Series expansions for the Ising spin glass in general dimension

\bibitem{adler91}
\bibinfo{author}{\bibfnamefont{L.}~\bibnamefont{Klein}},
\bibinfo{author}{\bibfnamefont{J.}~\bibnamefont{Adler}},
  \bibinfo{author}{\bibfnamefont{A.}~\bibnamefont{Aharony}},
  \bibinfo{author}{\bibfnamefont{A.~B.} \bibnamefont{Harris}},
   \bibnamefont{and}
  \bibinfo{author}{\bibfnamefont{Y.}~\bibnamefont{Meir}},
  \emph{\bibinfo{title}{Series expansions for the Ising spin glass in 
  general dimension}\/},
  \bibinfo{journal}{Phys. Rev. B} \textbf{\bibinfo{volume}{43}},
  \bibinfo{pages}{11249} (\bibinfo{year}{1991}).



\bibitem{Calabrese1}
P. Calabrese, V. Mart\'\i n-Mayor, A. Pelissetto, and E. Vicari,
{\em Three-dimensional randomly dilute Ising model: Monte Carlo results\/},
Phys. Rev. E {\bf 68}, 036136 (2003).


\bibitem{balles98}
H.~G. Ballesteros, L.~A. Fern{\'a}ndez, V.~Mart{\'\i}n-Mayor,
A.~Mu{\~n}oz~Sudupe, G.~Parisi, and J.~J. Ruiz-Lorenzo, {\em Critical
exponents of the three-dimensional diluted {I}sing model\/},
Phys. Rev. B {\bf 58}, 2740 (1998).
%%  --2747 [cond-mat/9802273].



\bibitem{Pelissetto1}
A. Pelissetto and E. Vicari,
{\em Randomly dilute spin models: A six-loop field-theoretic study\/},
Phys. Rev. B {\bf 62}, 6393 (2000).



%Random Ising model in three dimensions: theory, experiment and simulation 
% - a difficult coexistence

\bibitem{MC_here}
B.~Berche, P.-E.~Berche, C.~Chatelain, and W.~Janke, 
{\em Random Ising model in three dimensions: theory, experiment and simulation 
 -- a difficult coexistence\/},
preprint \eprint{cond-mat/0411255}.

\bibitem{kappler}
W.~Janke and S.~Kappler,
{\em Simulation of 3D $q$-state Potts models with the multibondic 
algorithm\/}, unpublished.



%Random-bond Potts models on hypercubic lattices:
%                          high-temperature series expansions",

\bibitem{hj:npb2002}
\bibinfo{author}{\bibfnamefont{M.}~\bibnamefont{Hellmund}} \bibnamefont{and}
  \bibinfo{author}{\bibfnamefont{W.}~\bibnamefont{Janke}},
  \emph{\bibinfo{title}{Random-bond Potts models on hypercubic lattices:
  High-temperature series expansions}\/},
  \bibinfo{journal}{Nucl. Phys. B (Proc. Suppl.)}
  \textbf{\bibinfo{volume}{106--107}}, \bibinfo{pages}{923}
  (\bibinfo{year}{2002}).



%"Star graph expansions for bond diluted Potts models"

\bibitem{hj:pre2003}
\bibinfo{author}{\bibfnamefont{M.}~\bibnamefont{Hellmund}} \bibnamefont{and}
  \bibinfo{author}{\bibfnamefont{W.}~\bibnamefont{Janke}},
  \emph{\bibinfo{title}{Star-graph expansions for bond diluted Potts models}\/},
  \bibinfo{journal}{Phys. Rev. E} \textbf{\bibinfo{volume}{67}},
  \bibinfo{pages}{026118} (\bibinfo{year}{2003}).



%Softening of first-order transition in 
%                  three-dimensions by quenched disorder
  
\bibitem{chat01a}
\bibinfo{author}{\bibfnamefont{C.}~\bibnamefont{Chatelain}},
  \bibinfo{author}{\bibfnamefont{B.}~\bibnamefont{Berche}},
  \bibinfo{author}{\bibfnamefont{W.}~\bibnamefont{Janke}}, \bibnamefont{and}
  \bibinfo{author}{\bibfnamefont{P.~E.} \bibnamefont{Berche}},
  \emph{\bibinfo{title}{Softening of first-order transition in
  three-dimensions by quenched disorder}\/},
  \bibinfo{journal}{Phys. Rev. E} \textbf{\bibinfo{volume}{64}},
  \bibinfo{pages}{036120} (\bibinfo{year}{2001}).


%Two-dimensional Ising-like systems: Corrections to scaling
%                 in the Klauder and double-Gaussian models

\bibitem{barma85}
\bibinfo{author}{\bibfnamefont{M.}~\bibnamefont{Barma}} \bibnamefont{and}
  \bibinfo{author}{\bibfnamefont{M.~E.} \bibnamefont{Fisher}},
  \emph{\bibinfo{series}{Two-dimensional Ising-like systems: Corrections to scaling
  in the Klauder and double-Gaussian models}\/},
  \bibinfo{journal}{Phys. Rev. B} \textbf{\bibinfo{volume}{31}},
  \bibinfo{pages}{5954} (\bibinfo{year}{1985}).

\end{thebibliography}
\bibliographystyle{apsrev}

\end{document}